\documentclass[aps,preprint,preprintnumbers,nofootinbib,showpacs]{revtex4}
\usepackage{graphicx,color,amsmath,amsfonts}

\def\fan{{f_A^{(n)}}}

\newcommand{\lsim}{\mathrel{\mathop{\kern 0pt \rlap
  {\raise.2ex\hbox{$<$}}}
  \lower.9ex\hbox{\kern-.190em $\sim$}}}
\newcommand{\gsim}{\mathrel{\mathop{\kern 0pt \rlap
  {\raise.2ex\hbox{$>$}}}
  \lower.9ex\hbox{\kern-.190em $\sim$}}}
\newcommand{\beq}     {\begin{equation}}
\newcommand{\eeq}     {\end{equation}}
\newcommand{\bea}     {\begin{eqnarray}}
\newcommand{\eea}     {\end{eqnarray}}
\newcommand{\no}     {\nonumber}
\newcommand{\lm}      {\lambda}
\newcommand{\di}      { \mathrm{d }}
\newcommand{\es}      {\epsilon}
\newcommand{\gm}      {\gamma}

\newcommand{\sg}      {\sigma}
\newcommand{\Dt}      {\Delta}
\newcommand{\kp}      {\kappa}

\newcommand{\ee}{e^+ e^-}
\newcommand{\dmq}      {\Delta M_q}
\newcommand{\dmd}      {\Delta M_d}
\newcommand{\dms}      {\Delta M_s}
\newcommand{\bb}      {{b\bar{b}}}

\newcommand{\bmix}      {B_q^0-\bar{B}_q^0}
\newcommand{\bdsmix}      {B_{d,s}^0-\bar{B}^0_{d,s}}
\newcommand{\bdmix}      {B_d^0-\bar{B}_d^0}
\newcommand{\bsmix}      {B_s^0-\bar{B}_s^0}
\def\lsim{\ ^<\llap{$_\sim$}\ }
\def\gsim{\ ^>\llap{$_\sim$}\ }

\def\n{{(n)}}

\def\fan{{f_A^{(n)}}}
\newcommand{\gauge}   {SU(3)$_c\,\times\, $SU(2)$_L\,\times\, $SU(2)$_R\,\times\,$U(1)$_{B-L}\,$}
\newcommand{\man}     {{ m_A^{(n)} }}
\newcommand{\xan}     {{ x_A^{(n)} }}
\newcommand{\ban}     {{ \beta_A^{(n)} }}
\newcommand{\Z}{\mathbb{Z}_2}
\newcommand{\zz}{\Z \times \Z'}
\newcommand{\szz}{S^1/\mathbb{Z}_2\times \mathbb{Z}_2'}
\begin{document}

%\begin{center}
%  {\Huge\bf \today}
%\end{center}

\title{Constraint of $\bdsmix$ mixing
on warped extra-dimension model}
%on the Randall-Sundrum model with custodial isospin symmetry}

\author{Sanghyeon Chang$^a$\footnote{schang@cskim.yonsei.ac.kr},~~
C. S. Kim$^{a,b}$\footnote{cskim@yonsei.ac.kr}~~ and~~ Jeonghyeon
Song$^c$\footnote{jhsong@konkuk.ac.kr}}

\affiliation{$^a$Department of Physics, Yonsei University,
Seoul 120-749, Korea\\
$^b$IPPP, Dept. of Physics, Univ. of Durham, Durham DH1 3LE, England\\
$^c$Department of Physics, Konkuk University, Seoul 143-701, Korea}
\date{\today}

\begin{abstract}
\noindent Recent CDF measurement of the $\bsmix$ oscillation frequency %, $\dms$,
at the Tevatron imposes significant constraint on various models for
new physics. A warped extra-dimension model with custodial isospin
symmetry accommodates the $\bdsmix$ mixing at tree level mainly
through the Kaluza-Klein gluons. This is due to the misalignment
between the bulk gauge eigenstates  and the localized Yukawa
eigenstates of the bulk fermions. We adopt the universal 5D Yukawa
coupling model where all Yukawa couplings are of order one. The SM
fermion mass spectra and mixings are controlled by the bulk Dirac
mass parameters. With two versions of the hadronic parameter values
for $\hat B_{B_{d,s}} f^2_{B_{d,s}}$, we investigate the implication
of the observed $\bdsmix$ mixings on this model. The CP-violating
effects on the $B_d$ system is shown to provide very strong
constraint: The first Kaluza-Klein mass of a gluon ($M_{KK}$) has
its lower bound about 3.7 TeV with $1\sigma$ uncertainty.
\end{abstract}
 \maketitle

\section{Introduction}
\label{sec:Introduction}

Recently CDF collaboration \cite{CDF} has measured the oscillation
frequency of  $\bsmix$ mixing using abundant $B_s$ mesons at the
Tevatron\footnote{ $\Delta M_s$ has been also observed by D{\O}
collaboration\,\cite{D0}, as $17 \,{\rm ps}^{-1}< \Delta M_s <
21\,{\rm ps}^{-1}$ at the 90\% C.L..}. $\bdmix$
mixing had been also measured by BaBar and Belle experiments at
the $\ee$ $B$ factories\,\cite{HFAG}. The current experimental
results are\,\cite{HFAG,CDF}
\bea \label{exp}
\Delta M_d^{\rm exp}&=&
(0.507\pm 0.004)\,{\rm ps}^{-1}\,, \\ \no \Delta M_s^{\rm exp} &=&
\left[17.33^{+0.42}_{-0.21}({\rm
  stat})\pm 0.07({\rm syst})\right]\,{\rm ps}^{-1}\,.
\eea
The observed oscillation frequency of $\bmix$ mixing
($q \in \{d,s \}$) determines the mass difference of $\bmix$ states,
$\dmq \equiv M_{B_q}^{\rm heavy} - M_{B_q}^{\rm light}$.
%The ratio of two experimental quantities are
%\beq
%\frac{\Delta M_s^{\rm exp}}{\Delta M_d^{\rm exp}}
%= 34.18^{+1.12}_{-0.70}\,.
%\eeq
For the $\bmix$ transition
amplitude $M_{\bb}^q$ defined by $\langle B_q^0| {\cal H}^{\Delta B=2}_{\rm eff} | \bar
B_q^0\rangle = 2 M_{B_q} M_{\bb}^q$,
its absolute value is determined by
the mass difference:
\beq
\dmq = 2\,
|M_{\bb}^q|.
\eeq
and the CP-violating phase $\phi_q$
generates ``mixing-induced" CP violation:
\beq
\label{eq:phiq:definition}
\phi_q = {\rm arg} \left( M_{\bb}^q \right).
\eeq

As a flavor-changing neutral-current (FCNC) process,
$\bmix$ mixing is a very
sensitive probe for the new physics beyond the standard model (SM)
since the SM contributions occur only at loop level.
%Moreover the ratio of two oscillation frequencies
%of the $\bdmix$ and $\bsmix$
%can provide more profound probe
%for the hadronic uncertainties are reduced for the ratio
%as shall be discussed below.
If any new physics model make its tree-level contribution to $\bmix$
mixing, the model  would be strongly
constrained. In the literature, the constraints on
various new models by $\bmix$ mixing have been extensively discussed\,\cite{other}.

Many new models are theoretically
motivated by the gauge hierarchy problem. Among them, a warped
extra dimension model by Randall and Sundrum (RS1)\,\cite{Randall:1999ee}
has attracted great interest, which  solves
the gauge hierarchy problem with geometrical suppression of Planck
scale to TeV scale. The RS1 model has one extra spatial dimension of
a truncated AdS space, the orbifold of $\szz$.
The fixed point under
$\Z$ parity transformation is called the Planck (UV) brane and that
under $\Z'$ parity the TeV (IR) brane.
In the original RS1 model, the SM
fields are localized on the TeV brane
in order to avoid any conflict with  (most
of) experimental data\,\cite{Davoudiasl:1999tf,Chang:1999nh}.
Later a bulk SM has been widely
studied because the phenomenological aspects of
the localized field in the 5D theory depend sensitively on the
unknown UV physics while those of the bulk field do
not\,\cite{Davoudiasl:1999tf,Chang:1999nh,Grossman:1999ra,
Gherghetta:2000qt,Huber:2000fh,Huber:2000ie}.
In addition, setting SM
fermions in the bulk can explain the enormous mass hierarchy between
top quark and neutrino without introducing the hierarchical Yukawa couplings and/or
seesaw mechanism\,\cite{Grossman:1999ra,Gherghetta:2000qt}.
However, many new strongly
interacting particles emerge around the TeV scale.
The electroweak precision data (EWPD) put very
strong constraint
mainly due to the lack of SU(2) custodial symmetry
in the theory\,\cite{Kim:2002kk,Csaki:2002gy,Hewett:2002fe,Burdman:2002gr}.

Later the SU(2) custodial symmetry was introduced
to be consistent with EWPD.
One of the most interesting approaches is the one
suggested by Agashe \emph{et.$\,$al.}\,\cite{Agashe:2003zs}.
In this model, SU(2) custodial  symmetry
is induced from
AdS$_5$/CFT feature of bulk gauge symmetry of \gauge.
The Higgs boson field remains as an ingredient,
which is confined on the TeV brane to avoid another
hierarchy problem\,\cite{Chang:1999nh}:
Its vacuum expectation value (VEV)
generates the SM particle masses.
For example, a SM fermion mass is determined
by its 5D Yukawa coupling and the overlapping magnitude of the
fermion zero mode function on the TeV brane.
The Kaluza-Klein (KK) mode function of a bulk fermion is
controlled by its bulk Dirac mass parameter $c$.
Unfortunately there is no unique way to determine 5D Yukawa coupling and $c$'s.
One reasonable and attractive choice is
to assign universal 5D Yukawa couplings of order
one to all fermions\,\cite{Huber:2003tu,Chang:2005ya,Agashe:2004ay,Agashe:2004cp}.
Small masses as well as small
mixings of the SM fermions (zero modes)
are explained by suppressed zero mode functions with moderate values of $c$'s.
Experimental data of SM flavor structure determine the $c$'s
under some reasonable assumptions.

\begin{figure}[b!]
  \includegraphics[scale=1]{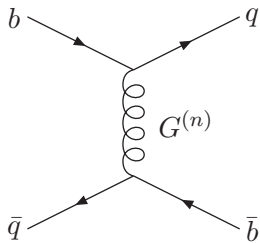}
   \caption{\label{fig}
Feynman diagram leading to $\bmix$ mixing in a warped extra
dimension model. $G^{(n)}$ is the $n$-th KK mode of a gluon.}
\end{figure}

The localized Yukawa couplings
generally mix the
zero modes of the bulk SM fermions of different generations,
while the bulk gauge interactions are flavor-diagonal.
This misalignment allows the
FCNC coupling at tree level,
as depicted in Fig.~\ref{fig}\,\cite{Agashe:2004ay,Agashe:2004cp}.
Since the coupling
strength of strong interaction is much larger than that of electroweak (EW)
interactions, the KK gluon contribution to $\Delta M_q$ is dominant.
These tree-level contributions to $\bdsmix$ mixings
can be {\it very sensitive probe to the custodial bulk {\rm RS1} model}.
This is the primary goal of the paper.

In Ref.~\cite{Agashe:2004cp}, a general argument
on the $\Delta F=2$ FCNC processes in this model was
discussed and the size of the new physics contribution was roughly estimated.
With the new experimental results on the $\bsmix$ mixing,
more comprehensive and detailed study
on this topic is worthwhile.
In this paper, we present the full formalism including general
complex phases in the left- and right-handed mixing matrices.
As shall be shown, the presence of complex phases is crucial
especially when we adopt a certain SM calculation of $\Dt M_q$.
Even though there is no prior knowledge on the value of the SM phase
$\phi^{\rm SM}_{d(s)}$, the approximate value of $\phi^{\rm SM}_d$ can be
indirectly deduced by combining  the value of $|V_{ub}|$ measured
from inclusive and/or exclusive tree level $b\rightarrow ul\nu$ decays and the
unitary phase angle $\gamma$ of the SM from the interference between
$b\rightarrow c$ and $b\rightarrow u$ transitions to $B\rightarrow DK^{(*)}$.
And we can get the constraints on the new physics
CP-violating phase $\phi^{\rm NP}_d$
by comparing the direct measurements of the CP phase from
$B\rightarrow J/\psi K_s$ decay~\cite{Agashe:2005hk,UTfit,Ball:EPJC}.
We will also examine how sensitive the new physics contribution to $\bmix$ mixing
is to the bulk fermion mass parameter.

The organization of the paper is as follows.  In the next section, we
briefly review the warped extra dimensional model
with \gauge.
We also formulate the new physics contribution to $\bmix$ mixing.
In Sec.~\ref{sec:numerical}, we summarize the current SM calculations
for $\bmix$, and examine the parameter space where the model
is compatible with the observed $\bdsmix$ mixings
and the CP phase $\phi_d$.
We conclude in Sec. \ref{sec:conclusions}.

\section{The warped extra-dimension model with custodial symmetry on TeV brane}
\label{sec:RS}

The basic set-up of the model is the same
as that of
references\,\cite{Agashe:2004ay,Agashe:2004cp,Huber:2003tu, Moreau:2006np, Chang:2005ya}.
We consider \gauge  gauge theory
in a five-dimensional warped space-time with the metric of
\begin{equation}
ds^2= e^{-2\sigma(y)}(dt^2-d\vec{x}^2) - dy^2,
\end{equation}
where $y$ is the fifth dimension coordinate and $\sigma(y) = k|y|$
with $k$ at the Planck scale. The theory is compactified on the $\szz$
orbifold, which is a circle (with radius $r_c$) compactified by
 two reflection
symmetries under $\Z \!:\! y \to -y$ and $\Z' \! : \! y'(=y-\pi r_c/2) \to -y'$.
In what follows, we denote $\Z$ parity by $P$, and $\Z'$ parity by
$P'$. Often conformal coordinate $ z\equiv e^{\sigma(y)}/k$ is more
useful with the metric of
\begin{equation}
ds^2= \frac{1}{(kz)^2}(dt^2-dx^2 - dz^2).
\end{equation}
The orbifold confines the fifth dimension $y \in [0,L(\equiv \pi
r_c/2)]$ or $z \in [1/k, 1/(e^{-kL}k)]$. With moderate value of $kL
\approx 35$, the IR cut-off $T\equiv e^{-kL}k$ can be at the TeV scale
so that the
gauge hierarchy problem is solved:
\beq T \equiv\es\, k \sim {\rm TeV} \quad
{\rm with}~ \es \equiv e^{-kL} \ll 1.
\eeq There are two fixed points in
the orbifold of $\szz$, the $\Z$-fixed point at $y=0$ ($z=1/k$)
called the Planck brane and the $Z'$-fixed point at $y=L$ ($z=1/T$)
called the TeV brane.

Among the bulk \gauge
gauge symmetry,
SU(2)$_R$ symmetry is broken by orbifold boundary conditions on
the Planck brane to U(1)$_R$:
We impose $(PP')=(-+)$ for
$\widetilde{W}_R^{1,2}$. Note that the TeV brane is  SU(2)$_R$
symmetric. The U(1)$_R\, \times\,$U(1)$_{B-L} $ is
spontaneously broken into U(1)$_Y$  on the Planck brane. Finally the EW symmetry
breaking of SU(2)$_L\, \times\,$ U(1)$_Y$ is triggered by the VEV of the
Higgs field localized on the TeV brane. The SM gauge
field is a zero mode of a bulk gauge field with $(++)$ parity.

A five-dimensional gauge field $A^M(x,z)$ (dimension 3/2) is
expanded in terms of KK modes:
\begin{eqnarray}
\label{eq:KKexpA} A_{\nu} (x,z) = \sqrt{k} \sum_n A_\nu^{(n)}(x)
\fan (z),
\end{eqnarray}
where the mode function $\fan(z)$ is dimensionless. The zero mode
function is
\beq \label{eq:fa0}
 f_A^{(0)} = \frac{1}{\sqrt{k L}},
\eeq
and the $n$-th ($n>0$) mode  function is
\begin{eqnarray}
\label{eq:fAn}
\fan(z)=  \frac{T z}{N_A^\n} \left[ J_1(\man z) +
\beta_A^\n Y_1(\man z) \right].
\end{eqnarray}
The following double constraints on $\beta_A^\n$ determines the KK
masses for the gauge bosons:
\beq \beta_A^\n = -
\frac{J_0(\man/T)}{Y_0(\man/T)} = -\frac{J_0(\man/k)}{Y_0(\man/k)}
\,.
\eeq
The normalization constant
is
\beq
N_A^{(n)}=  \left| \frac{\left[ J_1(\xan) + \ban Y_1(\xan)
\right]^2}{2}- 2\left( \frac{ \ban}{\pi \xan} \right)^2
\right|^{1/2}\,.
\eeq

The five-dimensional action of a bulk fermion $\Psi(x^\mu,y)$ is
\begin{eqnarray}
S_{\rm fermion}=\int d^4x \,\di y \left[ \overline{\hat\Psi}
e^\sigma i\gamma^\mu\partial_\mu \hat{\Psi} -\frac{1}{2}
\overline{\hat\Psi} \gamma_5\partial_y \hat{\Psi}
+\frac{1}{2}(\partial_y \overline{\hat\Psi})\gamma_5 \hat\Psi
 + m_D \overline{\hat\Psi}\hat\Psi
 \right]\ ,
\end{eqnarray}
where $\hat{\Psi}\equiv e^{-2\sigma}\Psi$ and the 5D Dirac mass is
$m_D = c \, k \,{\rm sign} (y)$. The bulk fermion field is
decomposed in terms of KK modes as
\beq
\label{eq:KKexpansion:Psi}
\hat{\Psi}(x,z) = \sqrt{k} \sum_{n} \left[ \psi^\n_L(x) f_{L}^\n(z)
+ \psi^\n_R(x) f^\n_R(z) \right] \,,
\eeq
where $\psi^\n_{R,L}(x) =\dfrac{1\pm \gm^5}{2}\psi^\n$.
The Dirac mass parameter
$c$ determines the KK mass spectrum and mode functions.
Since $\zz$ parity of $\Psi_L$ is always opposite to that of
$\Psi_R$, the left-handed SM fermion is the zero mode of a 5D
fermion whose left-handed part has $(++)$ parity (and the
right-handed part has automatically $(--)$ parity). And the
right-handed SM fermion, a singlet under SU(2)$_L$, is the zero mode of
\emph{another} 5D fermion whose right-handed part has $(++)$ parity.
We have two
non-vanishing zero mode functions,
\begin{eqnarray}
\label{eq:fLR0} f_{L}^{(0)}(z,c) = \frac{(T z)^{-c}}{N_L^{(0)}},\ \
\ f_{R}^{(0)}(z,c) = \frac{(T z)^{c}}{N_R^{(0)}},
\end{eqnarray}
with the  normalization constants of \beq N^{(0)}_L=\es^{-c} \left|
\frac{1-\es^{2c-1}}{2 c-1} \right|^{1/2}, \quad N^{(0)}_R=\es^{-c}
\left| \frac{1-\es^{-2c-1}}{-2 c-1} \right|^{1/2} \,. \eeq
%In a warped extra dimensional model with \gauge gauge symmetry,
%the SM left-handed leptons and
%quarks belong to 5D SU(2)$_L$ doublets whose left-handed part has
%$(++)$ parity. The SM right-handed fermions belong to SU(2)$_R$
%doublets with opposite $(PP')$ parity assignment.
Since the
$\widetilde{W}_R^{1,2}$ fields with $(-+)$ parity couples two
elements of a SU(2)$_R$ doublet, we have an extra $(-+)$ parity fermion
in each SU(2)$_R$ doublet.
Focusing on the $\bmix$ mixing which involves
only the zero modes for fermions,
we consider the fermion fields which contains the $(++)$ parity only. Then
the whole SM quarks can be contained in the bulk doublets,
\beq Q_i =\left(
       \begin{array}{c}
         u_{iL}^{(++)} \\
         d_{iL}^{(++)} \\
       \end{array}
     \right),
     \quad
U_i =\left(
       \begin{array}{c}
         u_{iR}^{(++)} \\
         D_{iR}^{(-+)} \\
       \end{array}
     \right),
     \quad
U_i =\left(
      \begin{array}{c}
        U_{iR}^{(-+)} \\
        d_{iR}^{(++)} \\
      \end{array}
    \right),
\eeq
where $i$ is the generation index. Note that nine Dirac mass
parameters ($c_{Q_i}$, $c_{U_i}$, and $c_{D_i}$) determine the
zero-mode functions and KK mass spectra in the quark sector.

To generate the SM fermion masses, we use the
localized Higgs field on the TeV brane as the usual SM Higgs mechanism.
The Yukawa interaction  between bulk quarks and Higgs is\,\cite{Chang:2005vj}
\beq S_Y = -\int \di^4 x \, \di y \frac{\delta(y-L)}{T} \left[
\lm^u_{5ij}
\overline{\hat{u}}_{iR}(x,y)\tilde{H}^\dagger(x)\hat{Q}_{jL}(x,y) +
\lm^d_{5ij} \overline{\hat{d}}_{iR}H^\dagger\hat{Q}_{jL} +h.c.
\right],
\eeq
where $H=e^{- k L}\phi(x)$ is canonically normalized
Higgs filed, $\tilde{H} = i \tau_2 H^*$,
and $i,~j$ are the generation indices. The boundary
mass term is realized when the Higgs field develops the VEV
of $\langle H \rangle = v$. The SM mass matrices for up- and
down-type quarks are then, for $q=U,D$,
\beq
\label{eq:Mq}
M^q_{ij}= v \lm^q_{5ij}
\left. \frac{k}{T} f_R^{(0)}(z,c_{q_i}) f_L^{(0)}(z,c_{Q_j})
\right|_{z=1/T}.
\eeq

The mass matrix $M^{u,d}$ is diagonalized by bi-unitary
transformation:
\begin{equation}
U_{qR} M^q U_{qL}^\dagger=M^{q}_{(\rm diag)} ~~~~~ \mbox{ for } q=u,d.
\label{bioth}
\end{equation}
The mass eigenstates are
\beq
\label{eq:chi-psi-mix}
\chi_{qL}=U_{qL} \psi^{(0)}_{qL},\quad \chi_{qR}=U_{qR} \psi^{(0)}_{qR}.
\eeq
The
Cabibbo-Kobayashi-Maskawa (CKM) matrix \,\cite{CKM}
is defined as $V^{\rm CKM}= U_{uL}^\dagger U_{dL}$.
A natural choice for $U_{uL}$ and $U_{dL}$ is that both mixing
matrices have similar form of the CKM matrix.
This choice of mixing is reasonable since the $u_L$ and $d_L$,
which belong to the same SU(2)$_L$ doublet, have
the same bulk mass. We parameterize
\begin{equation}
\left( U_{qL} \right)_{ij} = \kappa_{ij} V^{\rm CKM}_{ij},
 \label{mixing-quark}
\end{equation}
where $\kappa_{ij}$'s are complex parameters of order one.
To avoid
order changing during the diagonalization of matrix,
$\kp^2$ should be greater than $\sin\theta_c$
($\theta_c$ is the Cabibbo angle) and smaller than $1/\sin\theta_c$.
Therefore, we assume $|\kappa_{ij}| \in [1/\sqrt{2},2]$.

In the SM, the huge mass difference between electron and top quark
is explained by hierarchical Yukawa couplings.
Even though the  SM fermion mass in this model is also generated through the VEV of
the localized Higgs field, the mass hierarchy can be attributed to
different overlapping probability (by controlling $c$'s) of the
zero-mode function on the TeV brane. The SM
fermion mass spectra have been studied for various values of the
Dirac mass parameter $c$'s, and found that the large SM fermion mass
hierarchy can be explained without introducing large hierarchy in
the model parameters\,\cite{Agashe:2004ay, Agashe:2004cp,
Huber:2003tu, Moreau:2006np, Chang:2005ya}.

The structure of Yukawa couplings is arbitrary in this model.
One popular choice is to assume
that all of the 5D Yukawa couplings (to all flavors)
have almost universal strength of order one:
The fermion hierarchy is generated only by different
bulk Dirac mass parameters. Since the mass eigenvalues and CKM parameters are
empirically fixed, only unknowns are $\lambda^q_5$ of Eq.~(\ref{eq:Mq}) and
$\kappa_{ij}$ in Eq.~(\ref{mixing-quark}).
Since both parameters are all assumed to be of order
one, the numerical ambiguity of $\lambda^q_5$
can be absorbed into $\kappa$ by the redefinition of parameters.

In this set-up (where the 5D Yukawa couplings are
universal and the mixing matrices are CKM-like), the SM quark mass
spectrum is reproduced with the following Dirac mass
parameters\,\cite{Chang:2005ya}:
\begin{eqnarray}
\label{eq:c} c_{Q1} &\simeq &0.61,\ c_{Q2}  \simeq 0.56 ,\ c_{Q3}
\simeq 0.3^{+0.02}_{-0.04},
\\ \no
c_{D1} &\simeq& -0.66 ,\ c_{D2} \simeq -0.61 ,\ c_{D3} \simeq -0.56
,
\\ \no
c_{U1} &\simeq& -0.71 ,\ c_{U2} \simeq -0.53 ,\ 0\lsim c_{U3} \lsim
0.2.
\end{eqnarray}
As discussed before, the hierarchical SM mass spectrum
can be explained with moderate values of the model parameters
$c$'s.
At first glance, rather definite values of $c_{Q_{1,2}}$
seem unnatural, compared with $c_{Q_3} \in [0.26,0.32]$.
This is due to the behavior of the zero mode function, defined in Eq.~(\ref{eq:fLR0}),
in the fermion mass matrix of Eq.~(\ref{eq:Mq}).
For $c>0.5$, the $f^{(0)}_L$ is very sensitive to the change of $c$,
leading to strong constraint on $c$ from the fermion mass hierarchy:
The values of $c_{Q_{1,2}}$ are practically determined by the
SM quark mass hierarchy.
For $c<0.5$, however, the zero mode function does not change that much.
The value of $c_{Q_3}$ has some range, though small if we consider EW precision data and
Yukawa coupling around
$[1/\sqrt{2},2]$\,\cite{Agashe:2003zs}.

The 5D action for the gauge interaction of a fermion is
\beq
\label{eq:SG}
S_G =
\int \di^4 x \, \di y \sqrt{G} \frac{g_5}{\sqrt{k}} \bar{\Psi}(x,y)i
\gm^\mu A_\mu(x,y) \Psi(x,y),
\eeq
where $g_5$ is the 5D
dimensionless gauge coupling, and $\Psi = Q_{iL}, u_{iR}, d_{iR}$.
For the $\bmix$ mixing, only the zero modes of $Q_{iL}$ and
$ d_{iR}$ are relevant. With the preferred values of $c_i$'s in Eq.~(\ref{eq:c}),
the zero modes of $Q_{iL}$ are
dominant over those of $d_{iR}$.
Therefore, we take the contributions of the zero modes of $Q_{iL}$ only.
Substituting Eqs.~(\ref{eq:fa0}), (\ref{eq:fAn}) and (\ref{eq:fLR0})
into Eq.~(\ref{eq:SG})
leads to the four-dimensional gauge couplings, defined by
\beq
\mathcal{L}_{4D} =
g_{SM} \bar{\psi}_{jL}^{(0)} i \gm^\mu
\psi_{jL}^{(0)} A_\mu{(0)}
+g_{SM} \sum_{n=1}^\infty
\hat{g}_j^\n(c_{Q_i})\, \bar{\psi}_{jL}^{(0)}i \gm^\mu
\psi_{jL}^{(0)} A^\n_\mu,
\eeq
where $g_{SM} = g_5/\sqrt{kL}$ and
\beq
\hat{g}^\n_i(c_{Q_i}) = \sqrt{k L} \int \di z k \left[
f_L^{(0)}(z,c_{Q_i}) \right]^2 f_A^\n(z)\,.
\eeq
With the preferred
Dirac mass parameters in
Eq.~(\ref{eq:c}),
we have the following hierarchy in
$\hat{g}^\n_i$'s:
\beq
\label{eq:dominant-g}
\hat{g}^{(n)}_1 \sim
\lambda^2 \hat{g}^{(n)}_3, \quad \hat{g}^{(n)}_2 \sim \lm^2
\hat{g}^{(n)}_3,
\eeq
where $\lambda=\sin\theta_c\simeq 0.22$.
Thus, $\hat{g}^\n_3$ is dominant,
of which the value is determined by $c_{Q_3}$:
\beq
\hat{g}^{(1)}_3 = \left\{
                    \begin{array}{ll}
                      1.97 & \quad \hbox{for }c_{Q_3} = 0.3 \;; \\
                      1.80 & \quad\hbox{for }c_{Q_3} = 0.3 + 0.02\;; \\
                      2.31 & \quad\hbox{for }c_{Q_3} = 0.3 -0.04\;.
                    \end{array}
                  \right.
\eeq

The localized Yukawa interaction causes the mixing between the gauge
eigenstates and the mass eigenstates as in
Eq.~(\ref{eq:chi-psi-mix}).
The $n$-th KK  gauge interaction among
down-type quark mass eigenstates is
\beq
\mathcal{L}_{4D} \supset
g_{SM} \sum_{i,j=1}^3 \sum_{n=1}^\infty K^\n_{ij}
\,\bar{d}_{iL}^{(0)} i \gm^\mu d_{jL}^{(0)} A^\n_\mu,
\eeq
where
$d_{iL}$ is the mass eigenstate, and
\beq \label{eq:Kdef} K^\n
_{ij}= \sum_{k=1}^3 \big( U_{dL} \big)_{ik}
\,\hat{g}^\n_{k}(c_{Q_k}) \left( U_{dL}^\dagger \right)_{kj} \,.
\eeq
Note that if all of the $c_{Q_i}$ are the same for three
generations so that $\hat{g}^\n_{k}(c_{Q_k})$ is common, the
$K^\n_{ij}$ is proportional to $\delta_{ij}$:
No generation mixing and
thus no contribution to $\bmix$ mixing occur.

Through the $n$-th KK mode of a gluon exchange, $\bmix$ mixing amplitude is
proportional to
\begin{equation}
M^{q(n)}_{b\bar{b}}\propto \left(K^{(n)}_{q3}\right)^2 \simeq \left[
\kappa_{33} \kappa_{q3} V^{ *}_{tb} V^{}_{tq} \hat{g}^\n_3(c_{Q_3})
\right]^2,
\end{equation}
where  for the second equality we have included only the dominant
$\hat{g}_3^\n$ as in Eq. (\ref{eq:dominant-g}).

\section{The effects on $\bmix$ mixing}
\label{sec:numerical}

In the bulk RS model,
the $\bmix$ mixing is due to the SM box diagrams and the RS KK gluons:
\beq
M_{\bb}^{q} = M_{\bb}^{q, {\rm SM}}
\left( 1 +\frac{M_{\bb}^{q,{\rm RS}}}{M_{\bb}^{q, {\rm SM}} } \right).
\eeq
We parameterize, following the notation in Ref.~\cite{Ball:EPJC},
the new physics effect by
\beq
r_q e^{i \sigma_q} =\frac{M_{\bb}^{q,{\rm RS}}}{M_{\bb}^{q, {\rm SM}} },
\eeq
where $r_q \geq 0 $ and $\sigma_q$ is real.
The $r_q$ and $\sg_q$ are constrained by
the experimental result for $\Dt M_q$ and the theoretical
calculation of the $\Dt M _q^{\rm SM}$,
of which the ratio is defined by $\rho_q$:
\beq
\rho_q \equiv
\left|
\frac{\Dt M_q}{ \Dt M_q^{\rm SM}}
\right|=\left|\frac{M_{\bb}^{q, {\rm SM}}+M_{\bb}^{q,{\rm
RS}}}{M_{\bb}^{q, {\rm SM}}} \right|
=\sqrt{1+ 2 r_q \cos \sigma_q
+r_q^2}.
\eeq

The CP-violating phase $\phi_d$ can be divided into the phase from the SM
and that from New Physic (NP) contributions.
\beq
\phi_q = \phi_q^{\rm SM} + \phi_q^{\rm NP} =\phi_q^{\rm SM}
+{\rm arg } (1+ r_q e^{i\sigma_q}).
\eeq
The NP phase $\phi_q^{\rm NP} $ is determined by
$r_q$ and $\sigma_q$.
Therefore, the observation of $\phi_q^{\rm NP}$
can constrain $r_q$ and $\sg_q$, independently of $\rho_q$,
through the following relations:
\bea
\label{eq:sin}
\sin \phi_q^{\rm NP} &=&
\frac{r_q \sin\sigma_q}{ \sqrt{1+ 2r_q \cos\sg_q + r_q^2} }\,,
\\ \label{eq:cos}
\cos \phi_q^{\rm NP} &=&
\frac{1+ r_q \cos\sigma_q}{ \sqrt{1+ 2r_q \cos\sg_q + r_q^2} }\,.
\eea

The SM contribution is poorly known mainly due to
non-perturbative nature of the input hadronic parameters
even though experiments have measured $\dmd$ and $\dms$ with high precision.
$M_{12}^q$ within the SM is
\begin{equation}\label{M12SM}
M_{\bb}^{q,{\rm SM}} = \frac{G_{\rm F}^2
m_W^2}{12\pi^2}M_{B_q}\hat{\eta}^{B} \hat
B_{B_q}f_{B_q}^2(V_{tq}^\ast V_{tb})^2 S_0(x_t)\,,
\end{equation}
where $x_t = m_{top}^2/m_W^2$, and $S_0$ is an ``Inami--Lim" function\,\cite{IL}.
For CKM parameters, we used
$|V_{td}^* V_{tb}| = (8.6 \pm 1.3) \times 10^{-3}$
and $|V_{ts}^* V_{tb}| = (41.3 \pm 0.7) \times
10^{-3}$\,\cite{UTfit,Ball:EPJC}.
Common quantities for
both $B_d$ and $B_s$ system are $m_W$ and a
short-distance QCD correction $\hat{\eta}^{B}=0.552$\,\cite{Buras}.
Flavor dependent and non-perturbative quantities are the bag
parameter $\hat B_{B_q}$ and the decay constant $f_{B_q}$.
$r_q e^{i \sigma_q}$ in this model becomes
\beq
\label{eq:ratio}
r_q e^{i \sigma_q} \equiv \frac{M^{q,{\rm RS}}_{\bb}}{M^{q,{\rm SM}}_{\bb}}
 =
 \frac{16 \pi^2}{N_C} \frac{8 g_s^2}{g^4 S_0(x_t)}m_W^2 \kappa_{33}^2 \kappa_{q3}^2
 \sum_{n=1}\left( \frac{\hat{g}_3^\n (c_{Q_3})}{m_A^\n} \right)^2\,,
\eeq
where $\sigma_q = 2 \,{\rm arg}(\kp_{3q})$.
$r_q$ represents the magnitude of the new physics effect,
and $\sigma_q$ is a new source of CP violation.
Note that the parametrization in Eq.\,(\ref{mixing-quark}) removes
the CKM factor in the ratio.

There are several estimates of the SM values for
$\dmq^{\rm SM}$. We use the following two results
for the input hadronic parameters $\hat B_{B_{d,s}} f^2_{B_{d,s}}$.
The first one is from the most recent (unquenched) simulation by  JLQCD
collaboration\,\cite{JLQCD},
with non-relativistic $b$ quark and two flavors of dynamical light quarks.
The second one is from combined results, denoted by (HP+JL)QCD:
Lacking any direct calculation of $\hat B_{B_q}$ with three dynamical flavors,
it has been suggested to combine the results of
$f_{B_q}$ from HPQCD
collaboration\,\cite{HPQCD} with that of  $\hat B_{B_q}$ from
JLQCD.
Two numerical results are
\bea
\label{eq:DMd:SM}
\dmd^{\rm SM} &=&\left\{
                   \begin{array}{ll}
                     0.52 \pm 0.17 ^{-0.09}_{+0.13} ~{\rm ps}^{-1}, & \quad \hbox{for  JLQCD}\,;  \\
                     0.69 \pm 0.13 \pm 0.08~{\rm ps}^{-1}, & \quad \hbox{for  (HP+JL)QCD}\,,
                   \end{array}
                 \right.
\\ \label{eq:DMs:SM}
\dms^{\rm SM} &=& \left\{
                   \begin{array}{ll}
                     16.1\pm 2.8 ~{\rm ps}^{-1}, & \quad \hbox{for  JLQCD}\,;  \\
                    23.4 \pm 3.8 ~{\rm ps}^{-1}, & \quad \hbox{for  (HP+JL)QCD}\,,
                   \end{array}
                 \right.
\eea
where the first error in Eq.(\ref{eq:DMd:SM}) is from
the uncertainties of the CKM angle $\gamma$ and $R_b$,
while the second one is from those of $f_{B_d} \hat{B}_{B_d}^{1/2}$.

From the relation of $\Dt
M_q^{\rm RS}  = \Dt M_q^{\rm SM} \rho_q $ in this model,
we compute $\rho_q$ by using the experimental values and the SM values for $\Dt M_{d,s}$
as
\bea
\label{eq:DM:SM}
\rho_d &=&\left\{
                   \begin{array}{ll}
                     0.97 \pm 0.33 ^{-0.17}_{+0.26} , & \quad \hbox{for  JLQCD}\,;  \\
                     0.75 \pm 0.25 \pm 0.16, & \quad \hbox{for  (HP+JL)QCD}\,,
                   \end{array}
                 \right.
\\ \no
\rho_s &=& \left\{
                   \begin{array}{ll}
                     1.08^{+0.03}_{-0.01} ({\rm exp})\pm 0.19 ~({\rm th}), & \quad \hbox{for  JLQCD}\,;  \\
                    0.74^{+0.02}_{-0.01} ({\rm exp})\pm 0.18 ({\rm th}) , & \quad \hbox{for  (HP+JL)QCD}\,.
                   \end{array}
                 \right.
\,.
\eea

The CP-violating phase $\phi_d$
associated with $\bdmix$ is well measured,
of which the most recent average is \cite{HFAG}
\beq
\left( \sin \phi_d \right)_{c c \bar{s}} =
\sin (2 \beta + \phi_d^{\rm NP}) =0.687 \pm 0.032,
\eeq
where we have used $\phi_d^{\rm SM} = 2 \beta$.
The angle $\beta$ depends on two tree-level quantities of
$R_b$ and $\gamma$.
In spite of very large uncertainty in the angle $\gamma$,
the angle $\beta$ can be reasonably well constrained
since $\beta$ is very weakly affected by $\gamma$.
The new CP-violating phase is shown to be
\beq
\left. \phi_d^{\rm NP}  \right|_{\rm incl} = -(10.1 \pm 4.6)^\circ,
\quad
\left. \phi_d^{\rm NP}  \right|_{\rm excl} = -(2.5 \pm 8.0)^\circ.
\eeq
The detailed explanation for $\left. \phi_d^{\rm NP}  \right|_{\rm incl}$
and $\left. \phi_d^{\rm NP}  \right|_{\rm excl}$
is referred to Ref.~\cite{Ball:EPJC}.
Note that $\sin \sigma_d$ has the same sign with $\sin \phi_d^{\rm NP}$
through the relation in Eq.~(\ref{eq:sin}) with $r_q \geq 0$:
For $\left. \phi_d^{\rm NP}  \right|_{\rm incl}$,
therefore,
we explore the parameter $\sg_d$ in $[\pi, 2\pi]$
to satisfy Eq.~(\ref{eq:cos}).

%\FIGURE{\epsfig{file=dJLc3kp1_incl.eps,width=7cm}
%        \caption{Allowed parameter space of $(\sigma_d,~M_{KK})$}
%by $\rho_d$ and $\phi_d^{NP}|_{\rm incl}$. Three red (thin) lines
%satisfy the observed $\rho_d$ and three blue (thick) lines for
%$\phi_d^{\rm NP}|_{\rm incl}$, both with 1$\sigma$ uncertainty. We
%set $c_{Q_3}=0.3$ and $\kp = 1$, and use JLQCD hadronic input
%parameters.}%
%    \label{fig:ContourJL:incl}
%    }
\begin{figure}[t!]
\begin{center}
    \includegraphics[scale=0.7]{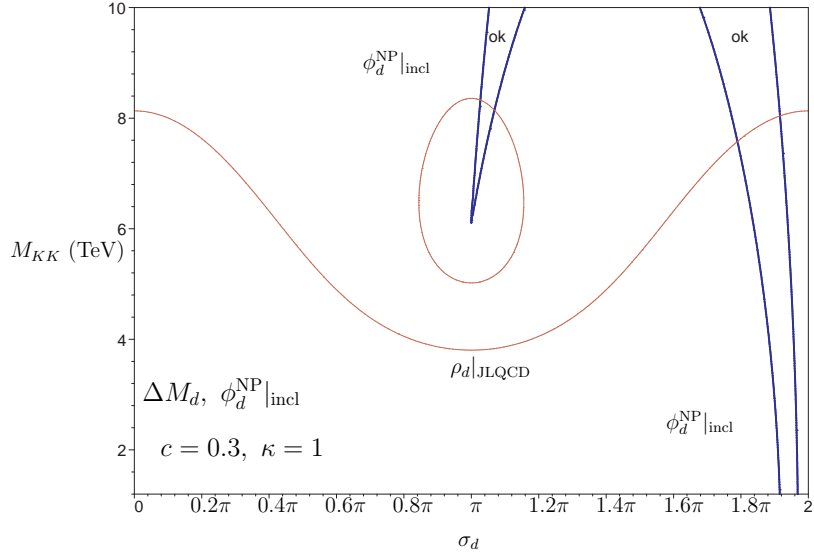}
    \end{center}
    \caption {Allowed parameter space of $(\sigma_d,~M_{KK})$
by $\rho_d$ and $\phi_d^{NP}|_{\rm incl}$. Two red (thin) lines
satisfy the observed $\rho_d$ and two blue (thick) lines for
$\phi_d^{\rm NP}|_{\rm incl}$, both with 1$\sigma$ uncertainty. We
set $c_{Q_3}=0.3$ and $\kp = 1$, and use JLQCD hadronic input
parameters.
    }
    \label{fig:ContourJL:incl}
\end{figure}

The new contribution depends on two parameters, $\sigma_d$ and
$M_{KK}$. Here $M_{KK}$ denotes the TeV scale mass
of the first KK mode of a gluon.
In Fig.~\ref{fig:ContourJL:incl}, we show in the parameter space of
$(\sigma_d,M_{KK})$ the contours for
$\rho_d$ (three red lines) and $\phi_d^{\rm NP}|_{\rm incl}$ (three blue lines)
with 1$\sg$ hadronic uncertainty.
With $c_{Q_3}=0.3$ and $ \kp=1$,
we use the JLQCD ones for the SM results.
The yellow region is allowed by two observation of
$\rho_d$ and $\phi_d^{\rm NP}|_{\rm incl}$.
Only with the $\rho_d$ constraint,
$M_{KK}$ can be as low as about 4 TeV and $\sigma_d$
in the whole range can be accepted.
The $\phi_d^{\rm NP}$ constraint is quite strong
to raise the allowed $M_{KK}$ above 8 TeV.
This heavy KK mode is practically impossible to probe at LHC.
In addition, the allowed region for $\sg_d$ is quite limited
to two small regions around $\pi$ and $1.8\pi$.

%\FIGURE{\epsfig{file=dJLc3kp1_excl.eps,width=8cm}
%        \caption{The same plots as in Fig.~\ref{fig:ContourJL:incl}
%    except for $\phi_d^{\rm NP}|_{\rm\bf excl}$.}%
%    \label{fig:ContourJL:excl}}
\begin{figure}[t!]
\begin{center}
    \includegraphics[scale=0.7]{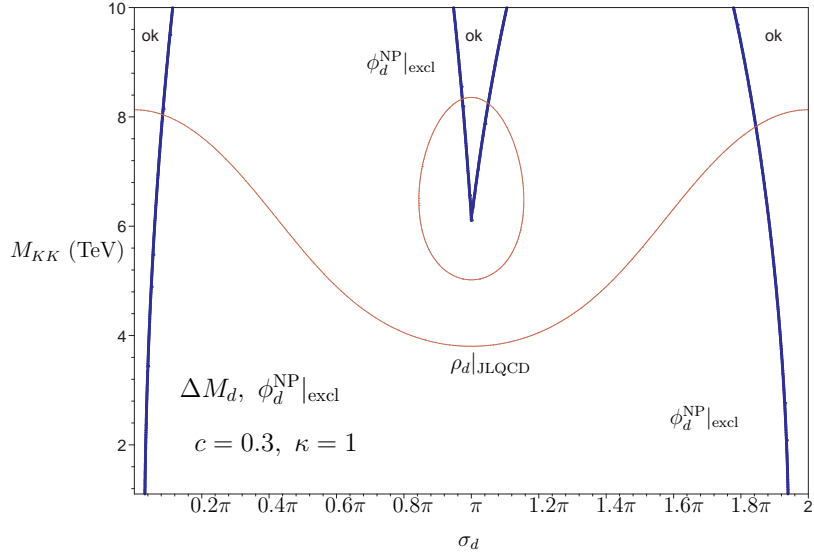}
    \end{center}
    \caption {The same plots as in Fig.~\ref{fig:ContourJL:incl}
    except for $\phi_d^{\rm NP}|_{\rm\bf excl}$.
    }
    \label{fig:ContourJL:excl}
\end{figure}

\begin{figure}[b!]
\begin{center}
    \includegraphics[scale=0.7]{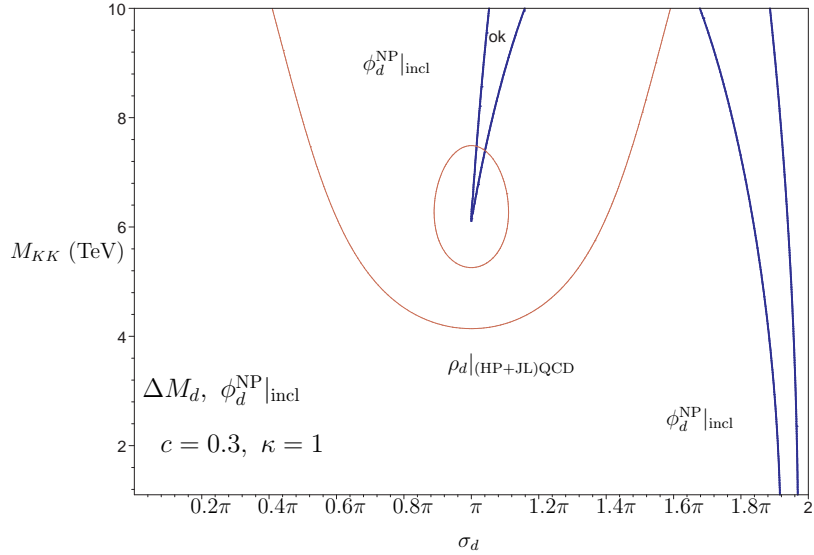}
    \end{center}
    \caption {The sample plot for (HP+JL)QCD hadronic inputs.
    }
    \label{fig:ContourHP:incl}
\end{figure}

Figure \ref{fig:ContourJL:excl} shows the same
plots but with different CP-violating
phase $\phi_d^{\rm NP}|_{\rm excl}$.
Large uncertainty in $\phi_d^{\rm NP}|_{\rm excl}$
does not lower the allowed $M_{KK}$ significantly.
In Fig.~\ref{fig:ContourHP:incl},
we present the same allowed parameter space but with different
hadronic input parameters, the (HP+JL)QCD one.
The $\rho_d$ constraint becomes stronger,
which narrows the allowed $\sigma_d$ region:
$\sg_d$ is limited around $[\pi, 1.2\pi]$.
The lowest allowed $M_{KK}$ is a little reduced, but not significantly.
It is still difficult to produce at LHC.

%\FIGURE{\epsfig{file=dJLc32kp1sq_incl.eps,width=5cm}
%        \caption {The same plots as in Fig.~\ref{fig:ContourJL:incl}
%but with $c_{Q_3}=0.32$ and $\kp=1/\sqrt{2}$.
%    }
%    \label{fig:ContourJL:min}}
\begin{figure}[t!]
\begin{center}
    \includegraphics[scale=0.7]{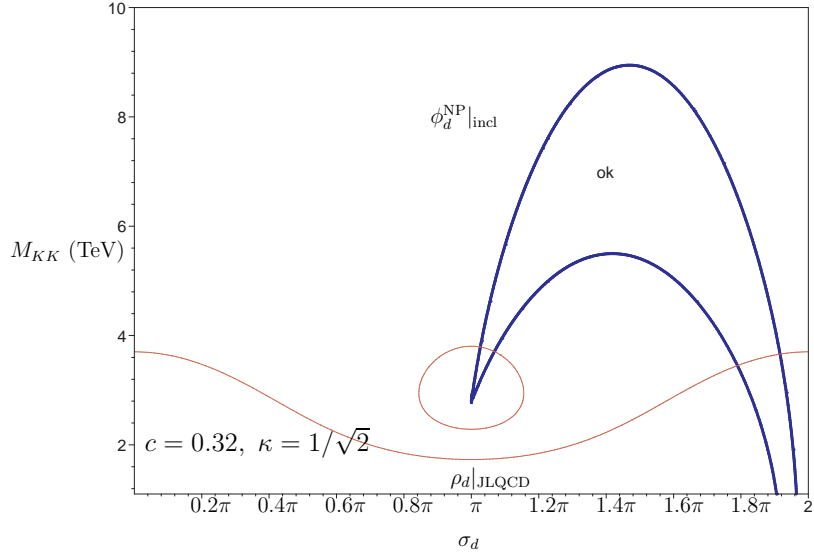}
    \end{center}
    \caption {The same plots as in Fig.~\ref{fig:ContourJL:incl}
but with $c_{Q_3}=0.32$ and $\kp=1/\sqrt{2}$.
    }
    \label{fig:ContourJL:min}
\end{figure}
In Fig.~\ref{fig:ContourJL:min},
we consider other values for $c_{q_3}$ and $\kp$
to see how much they affect the lowest allowed value for $M_{KK}$.
As can be seen from Eq.(\ref{eq:ratio}),
the smaller $\kp$ and $\hat{g}^{(1)}_3$
allow the lower $M_{KK}$ value.
In Fig.~\ref{fig:ContourJL:min}, we
present the allowed parameter space with
$c_{Q_3}=0.32$ and $\kp=1/\sqrt{2}$.
The allowed region of $\sg_d$ is significantly extended
into $[\pi,1.9\pi]$.
The allowed $M_{KK}$ is also remarkably reduced around 3.7 TeV so that
the first KK gauge boson can be marginally produced at LHC.
Note that we can choose even lower
$\kp<1/\sqrt{2}$ to allow lower $M_{KK}$. However, such a choice
may ruin the consistency of the bulk mass form in Eq.(\ref{eq:c}).

The CP-violating phase $\phi_s$, which
enters the $\bsmix$ mixing-induced CP-violation,
has not been constrained to this day.
The $\sigma_s$ measurement in the future experiment
is of great significance.
For example, the $A_{CP} (B_x \to \psi\phi)$
and the semi-leptonic asymmetry
$A^s_{SL}$ have very suppressed contribution from the SM.
Any significant measurement can indicate the new physics effect,
which the bulk Randall-Sundrum model under consideration can provide.

\begin{figure}[t!]
\begin{center}
    \includegraphics[scale=0.5]{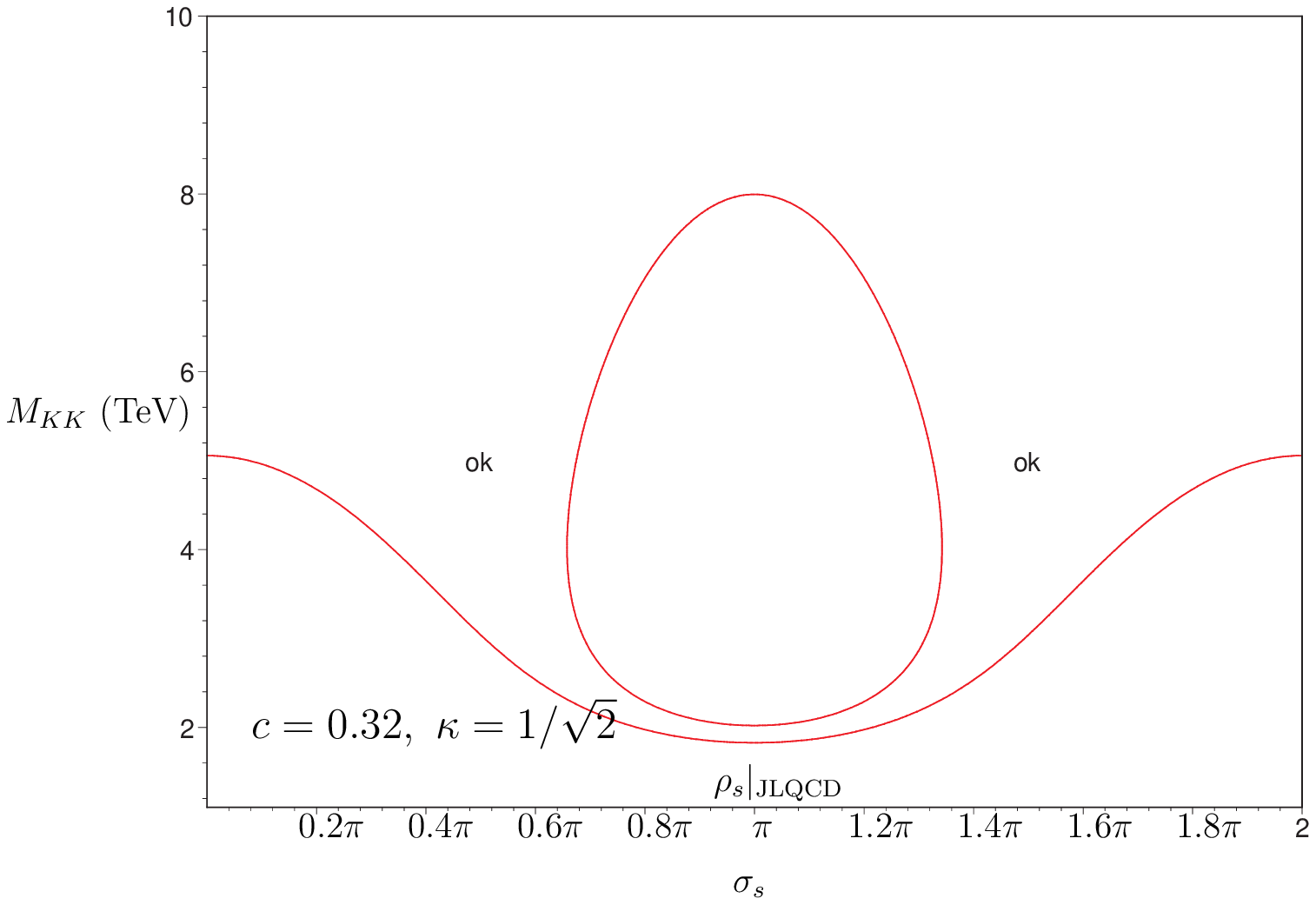}
 \includegraphics[scale=0.5]{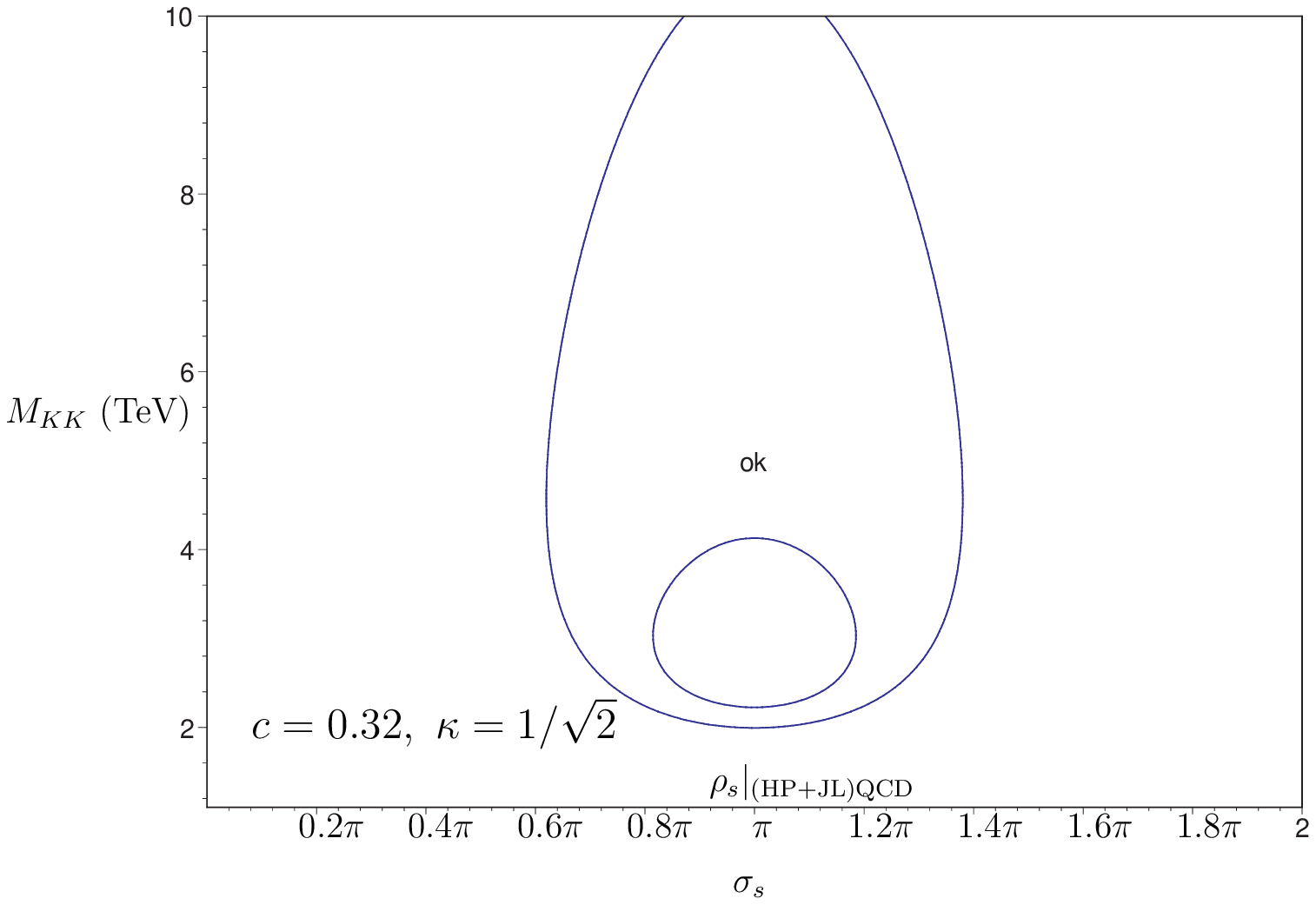}
    \end{center}
    \caption {The allowed parameter space of
$(\sg_s,~M_{KK})$
from the observed $\Dt M_s$.
We set $c_{Q_3}=0.32$ and $\kp=1/\sqrt{2}$
and used JLQCD and (HP+JL)QCD hadronic input parameters.
    }
    \label{fig:s}
\end{figure}

In Fig.~(\ref{fig:s}),
we plot the allowed region by the observed $\Dt M_s$.
We consider the $c_{Q_3}=0.32$ and $\kp=1/\sqrt{2}$ case
with both of JLQCD and (HP+JL)QCD hadronic input parameters.
Without any information on $\phi_s$,
there exist narrow but substantial region for $M_{KK} \simeq 2$ TeV.
However, this region is excluded by the observed $\phi_d$.

\section{Conclusions}
\label{sec:conclusions}

The warped extra dimensional model with custodial isospin
symmetry can contribute to $\bdsmix$ mixing at tree level, dominantly through
the Kaluza-Klein modes of  gluons.
This FCNC process at tree level originates from the
mixing between the gauge eigenstates and the mass eigenstates due to the flavor-mixing
Yukawa couplings localized on the TeV brane: Even
though the FCNC among the SM fermions (or
zero modes of the bulk fermion) is absent in the 5D gauge
interaction, the localized Yukawa couplings can mix the gauge
eigenstates. We assume the simplest set-up for the SM mass spectra
such that all the 5D Yukawa couplings are of the same order and the
mixing matrices have the same form as the CKM matrix. This
assumption almost fixes the bulk Dirac mass parameters $c_i$ for each
5D fermion. With the suggested $c_i$, we have calculated the new
contributions to the $\bdsmix$ mixings, and compare with the recent
experimental results. Main uncertainties are from
the hadronic input parameters. However, we found that the lower bound on $M_{KK}$
by the observed $\bmix$ mixing
is, irrespective to hadronic uncertainties, rather high above $\sim 3.7$ TeV.
The LHC can marginally produce the KK gluons.
The strongest constraint comes from the observation of the CP-violating phase $\phi_d$.

\acknowledgments
The work of C.S.K. was supported
in part by  CHEP-SRC Program,
in part by the Korea Research Foundation Grant funded by the
Korean Government (MOEHRD) No. KRF-2005-070-C00030.
The work of JS and SC are supported by the Korea Research Foundation Grant (KRF-2005-070-C00030).


\begin{thebibliography}{99}

\bibitem{CDF}G.~Gomez-Ceballos [CDF coll.], talk at FPCP 2006, {\tt
  http://fpcp2006.triumf.ca}.

\bibitem{D0}V.~M.~Abazov {\it et al.}  [D0 Collaboration],
  %``First direct two-sided bound on the B/s0 oscillation frequency,''
  arXiv:hep-ex/0603029.
  %%CITATION = HEP-EX 0603029;%%

\bibitem{HFAG}[Heavy Flavor Averaging Group (HFAG)],
  %``Averages of b-hadron properties at the end of 2005,''
  arXiv:hep-ex/0603003.
  %%CITATION = HEP-EX 0603003;%%



\bibitem{other}
M.~Carena, A.~Menon, R.~Noriega-Papaqui, A.~Szynkman and
C.~E.~M.~Wagner,
%   ``Constraints on B and Higgs physics in minimal low energy supersymmetric
%   models,''
  %
  arXiv:hep-ph/0603106;
  %%CITATION = HEP-PH 0603106;%%
M.~Ciuchini and L.~Silvestrini,
%   ``Upper bounds on SUSY contributions to b $\to$ s transitions from B/s -
%   anti-B/s mixing,''
  %
  arXiv:hep-ph/0603114;
  %%CITATION = HEP-PH 0603114;%%
L.~Velasco-Sevilla,
   %``Impact of Delta(m(B/s)) on the determination of the unitary triangle and
%   bounds on physics beyond the standard model,''
  %
  arXiv:hep-ph/0603115;
  %%CITATION = HEP-PH 0603115;%%
M.~Endo and S.~Mishima,
   %``Constraint on right-handed squark mixings from B/s - anti-B/s mass
%   difference,''
  %
  arXiv:hep-ph/0603251;
  %%CITATION = HEP-PH 0603251;%%
M.~Blanke, A.~J.~Buras, D.~Guadagnoli and C.~Tarantino,
   %``Minimal flavour violation waiting for precise measurements of Delta(M(s)),
%   $|$V(ub)$|$, gamma and B/s,d0 $\to$ mu+ mu-,''
  %
  arXiv:hep-ph/0604057.
Z.~Ligeti, M.~Papucci and G.~Perez,
   %``Implications of the measurement of the B/s0 - anti-B/s0 mass difference,''
  %
  arXiv:hep-ph/0604112;
  %%CITATION = HEP-PH 0604112;%%
J.~Foster, K.~i.~Okumura and L.~Roszkowski,
   %``New constraints on SUSY flavour mixing in light of recent measurements at
%   the Tevatron,''
  %
  arXiv:hep-ph/0604121;
%P.~Ball and R.~Fleischer,
%  %``Probing new physics through B mixing: Status, benchmarks and prospects,''
%  arXiv:hep-ph/0604249;
%  %%CITATION = HEP-PH 0604249;%%
Y.~Grossman, Y.~Nir and G.~Raz,
  %``Constraining the phase of B/s - anti-B/s mixing,''
  arXiv:hep-ph/0605028;
  %%CITATION = HEP-PH 0605028;%%
K.~Cheung, C.~W.~Chiang, N.~G.~Deshpande and J.~Jiang,
  %``Constraints on flavor-changing Z' models by B/s mixing, Z' production, and
  %B/s $\to$ mu+ mu-,''
  arXiv:hep-ph/0604223;
  %%CITATION = HEP-PH 0604223;%%
G.~Isidori and P.~Paradisi,
  %``Hints of large tan(beta) in flavour physics,''
  arXiv:hep-ph/0605012;
  %%CITATION = HEP-PH 0605012;%%
S.~Khalil,
  %``Supersymmetric contribution to the CP asymmetry of B $\to$ J/psi Phi in the
  %light of recent B/s - anti-B/s measurements,''
  arXiv:hep-ph/0605021;
  %%CITATION = HEP-PH 0605021;%%
A.~Datta,
  %``B/s mixing and new physics in hadronic b $\to$ s anti-q q transitions,''
  arXiv:hep-ph/0605039;
  %%CITATION = HEP-PH 0605039;%%
S.~Baek,
  %``B/s - anti-B/s mixing in the MSSM scenario with large flavor mixing in the
  %LL/RR sector,''
  arXiv:hep-ph/0605182;
  %%CITATION = HEP-PH 0605182;%%
X.~G.~He and G.~Valencia,
  %``anti-B/s B/s mixing constraints on FCNC and a non-universal Z',''
  arXiv:hep-ph/0605202;
  %%CITATION = HEP-PH 0605202;%%
R.~Arnowitt, B.~Dutta, B.~Hu and S.~Oh,
  %``B/s - anti-B/s mixing and its implication for b $\to$ s transitions in
  %supersymmetry,''
  arXiv:hep-ph/0606130;
  %%CITATION = HEP-PH 0606130;%%
S.~Baek, J.~H.~Jeon and C.~S.~Kim,
  %``B/s0 - anti-B/s0 mixing in leptophobic Z' model,''
  arXiv:hep-ph/0607113.
  %%CITATION = HEP-PH 0607113;%%



%G.~Buchalla, A.~J.~Buras and M.~E.~Lautenbacher,
%  %``Weak Decays Beyond Leading Logarithms,''
%  Rev.\ Mod.\ Phys.\  {\bf 68}, 1125 (1996)
%  [arXiv:hep-ph/9512380].
%  %%CITATION = HEP-PH 9512380;%%

%\cite{Randall:1999ee}
\bibitem{Randall:1999ee}
  L.~Randall and R.~Sundrum,
  %``A large mass hierarchy from a small extra dimension,''
  Phys.\ Rev.\ Lett.\  {\bf 83}, 3370 (1999).
  %%CITATION = HEP-PH 9905221;%%

%\cite{Davoudiasl:1999tf}
\bibitem{Davoudiasl:1999tf}
  H.~Davoudiasl, J.~L.~Hewett and T.~G.~Rizzo,
  %``Bulk gauge fields in the Randall-Sundrum model,''
  Phys.\ Lett.\ B {\bf 473}, 43 (2000).
  %%CITATION = HEP-PH 9911262;%%

%\cite{Chang:1999nh}
\bibitem{Chang:1999nh}
  S.~Chang, J.~Hisano, H.~Nakano, N.~Okada and M.~Yamaguchi,
  %``Bulk standard model in the Randall-Sundrum background,''
  Phys.\ Rev.\ D {\bf 62}, 084025 (2000).
  %%CITATION = HEP-PH 9912498;%%


\bibitem{Huber:2000fh}
S.~J.~Huber and Q.~Shafi,
%``Higgs mechanism and bulk gauge boson masses in the Randall-Sundrum  model,''
Phys.\ Rev.\ D {\bf 63}, 045010 (2001).
%%CITATION = HEP-PH 0005286;%%

%\cite{Huber:2000ie}
\bibitem{Huber:2000ie}
S.~J.~Huber and Q.~Shafi,
%``Fermion masses, mixings and proton decay in a Randall-Sundrum model,''
Phys.\ Lett.\ B {\bf 498}, 256 (2001).
%%CITATION = HEP-PH 0010195;%%

%\cite{Grossman:1999ra}
\bibitem{Grossman:1999ra}
  Y.~Grossman and M.~Neubert,
  %``Neutrino masses and mixings in non-factorizable geometry,''
  Phys.\ Lett.\ B {\bf 474}, 361 (2000).
  %%CITATION = HEP-PH 9912408;%%

%\cite{Gherghetta:2000qt}
\bibitem{Gherghetta:2000qt}
  T.~Gherghetta and A.~Pomarol,
  %``Bulk fields and supersymmetry in a slice of AdS,''
  Nucl.\ Phys.\ B {\bf 586}, 141 (2000).
  %%CITATION = HEP-PH 0003129;%%

%\cite{Kim:2002kk}
\bibitem{Kim:2002kk}
C.~S.~Kim, J.~D.~Kim and J.~Song,
%``Top quark Kaluza-Klein mode mixing in the Randall-Sundrum bulk standard
%model and B $\to$ X/s gamma,''
Phys.\ Rev.\ D {\bf 67}, 015001 (2003).
%%CITATION = HEP-PH 0204002;%%



%\cite{Csaki:2002gy}
\bibitem{Csaki:2002gy}
C.~Csaki, J.~Erlich and J.~Terning,
% ``The effective Lagrangian in the Randall-Sundrum model and electroweak
%physics,''
%
Phys.\ Rev.\ D {\bf 66}, 064021 (2002).
%%CITATION = HEP-PH 0203034;%%


%\cite{Hewett:2002fe}
\bibitem{Hewett:2002fe}
J.~L.~Hewett, F.~J.~Petriello and T.~G.~Rizzo,
% ``Precision measurements and fermion geography in the Randall-Sundrum  model
%revisited,''
%
JHEP {\bf 0209}, 030 (2002).
%%CITATION = HEP-PH 0203091;%%

%\cite{Burdman:2002gr}
\bibitem{Burdman:2002gr}
G.~Burdman,
%``Constraints on the bulk standard model in the Randall-Sundrum scenario,''
Phys.\ Rev.\ D {\bf 66}, 076003 (2002).
%%CITATION = HEP-PH 0205329;%%



%\cite{Agashe:2003zs}
\bibitem{Agashe:2003zs}
  K.~Agashe, A.~Delgado, M.~J.~May and R.~Sundrum,
  %``RS1, custodial isospin and precision tests,''
  JHEP {\bf 0308}, 050 (2003).
  %%CITATION = HEP-PH 0308036;%%
%\cite{Huber:2000fh}

%\cite{Huber:2003tu}
\bibitem{Huber:2003tu}
  S.~J.~Huber,
  %``Flavor violation and warped geometry,''
  Nucl.\ Phys.\ B {\bf 666}, 269 (2003).
  %%CITATION = HEP-PH 0303183;%%

%\cite{Chang:2005ya}
\bibitem{Chang:2005ya}
  S.~Chang, C.~S.~Kim and M.~Yamaguchi,
  %``Hierarchical mass structure of fermions in warped extra dimension,''
  Phys.\ Rev.\ D {\bf 73}, 033002 (2006).
  %%CITATION = HEP-PH 0511099;%%

%\cite{Agashe:2004ay}
\bibitem{Agashe:2004ay}
  K.~Agashe, G.~Perez and A.~Soni,
  %``B-factory signals for a warped extra dimension,''
  Phys.\ Rev.\ Lett.\  {\bf 93}, 201804 (2004).
  %%CITATION = HEP-PH 0406101;%%

%\cite{Agashe:2004cp}
\bibitem{Agashe:2004cp}
  K.~Agashe, G.~Perez and A.~Soni,
  %``Flavor structure of warped extra dimension models,''
  Phys.\ Rev.\ D {\bf 71}, 016002 (2005).
  %%CITATION = HEP-PH 0408134;%%


%\cite{Agashe:2005hk}
\bibitem{Agashe:2005hk}
  K.~Agashe, M.~Papucci, G.~Perez and D.~Pirjol,
  %``Next to minimal flavor violation,''
  arXiv:hep-ph/0509117.
  %%CITATION = HEP-PH 0509117;%%

\bibitem{UTfit}
%\cite{Bona:2006sa}
%\bibitem{Bona:2006sa}
  M.~Bona {\it et al.}  [UTfit Collaboration],
  %``The UTfit collaboration report on the unitarity triangle beyond the
  %standard model: Spring 2006,''
  Phys.\ Rev.\ Lett.\  {\bf 97}, 151803 (2006);
%  [arXiv:hep-ph/0605213].
  %%CITATION = HEP-PH 0605213;%%
A. Stocchi, private communication.

%\cite{Ball:2006xx}
\bibitem{Ball:EPJC}
  P.~Ball and R.~Fleischer,
  %``Probing new physics through B mixing: Status, benchmarks and prospects,''
  Eur.\ Phys.\ J.\ C {\bf 48}, 413 (2006)
  [arXiv:hep-ph/0604249].
  %%CITATION = HEP-PH 0604249;%%


%\cite{Moreau:2006np}
\bibitem{Moreau:2006np}
  G.~Moreau and J.~I.~Silva-Marcos,
  %``Flavour physics of the RS model with KK masses reachable at LHC,''
  JHEP {\bf 0603}, 090 (2006).
  %%CITATION = HEP-PH 0602155;%%


\bibitem{CKM}N.~Cabibbo,
  %``Unitary Symmetry And Leptonic Decays,''
 Phys.\ Rev.\ Lett.~{\bf 10} (1963) 531.
  %%CITATION = PRLTA,10,531;%%
M.~Kobayashi and T.~Maskawa,
%``CP Violation In The Renormalizable Theory Of Weak Interaction,''
 Prog.\ Theor.\ Phys.~{\bf 49} (1973) 652.
%%CITATION = PTPKA,49,652;%%

\bibitem{IL}
T.~Inami and C.~S.~Lim,
  Prog.\ Theor.\ Phys.\  {\bf 65}, 297 (1981)
  [Erratum-ibid.\  {\bf 65}, 1772 (1981)].
  %%CITATION = PTPKA,65,297;%%

\bibitem{Buras}
  G.~Buchalla, A.~J.~Buras and M.~E.~Lautenbacher,
  %``Weak Decays Beyond Leading Logarithms,''
  Rev.\ Mod.\ Phys.\  {\bf 68}, 1125 (1996).
  %%CITATION = HEP-PH 9512380;%%

\bibitem{JLQCD}S.~Aoki {\it et al.}\  [JLQCD coll.],
  %``B0 anti-B0 mixing in unquenched lattice QCD,''
  Phys.\ Rev.\ Lett. {\bf 91} (2003) 212001.
%  [arXiv:hep-ph/0307039].
  %%CITATION = HEP-PH 0307039;%%

\bibitem{HPQCD}A.~Gray {\it et al.}\  [HPQCD coll.],
  %``The B meson decay constant from unquenched lattice QCD,''
  Phys.\ Rev.\ Lett. {\bf 95} (2005) 212001.
%  [arXiv:hep-lat/0507015].
  %%CITATION = HEP-LAT 0507015;%%

%****************888

%\cite{Chang:2005vj}
\bibitem{Chang:2005vj}
  S.~Chang, S.~C.~Park and J.~Song,
  %``Kaluza-Klein masses of bulk fields with general boundary conditions in
  %AdS(5),''
  Phys.\ Rev.\ D {\bf 71}, 106004 (2005).
  %%CITATION = HEP-PH 0502029;%%











%---------------------------------

\end{thebibliography}
\end{document}